\documentclass{article}
\usepackage{cite}
\usepackage{indentfirst}

\title{With Great Power Comes Great Responsibility: Utilizing Privacy Technology for the Greater Bad}
\author{Sarah Cline and Jacob Aronoff}
\date{12/14/19}

\begin{document}

\maketitle

\section{Introduction}
As people across the world become increasingly aware of how their privacy is compromised in this digital era, the field of Privacy Enhancing Technologies, or PETs, has boomed. The first workshop on Privacy Enhancing Technology was in 2000, then called the "Workshop on Design Issues in Anonymity and Unobservability" \cite{petsym}. By 2007, the workshop had ballooned into a full symposium. In 2015, the first issue of the Proceedings on Privacy Enhancing Technologies journal was published \cite{privenhanctech}. This year in 2019, there were 4 volumes of Proceedings on Privacy Enhancing Technologies published, containing a total of 66 papers. Of these papers, we identified 14 which specifically describe PETs that have been newly developed or utilized in a new way. We focused on 3 papers that seemed to have widespread use cases. Some of these use cases, however, include criminal activity. While heavily focused on adversarial actions and capability, much of security research does not focus on or consider the social science and motivations behind bad actors. We believe that this is a critical factor to consider when working on the problem of cybercrime. In this paper, we will analyze some of the cutting edge PETs and the potential risks associated with them from the lenses of computer science and criminal justice. We argue that the continued development of privacy enhancing technology contributes to the global rise of cybercrime. 

\section{Crime}
The field of Criminal Justice is built on theories. Researchers will examine crime data and try to find patterns based off of criminology, sociology, psychology, and even economics. These theories focus on different elements of crime, such as the different types of crime, the different types of offenders, how to deal with crime after it happens, and how to deal with crime before it happens.

The idea of deterrence is a continuously researched topic in the field of Criminal Justice that focuses on trying to deal with crime before it happens. Researchers hope to understand how we can de-incentivise and prevent criminal behaviors by making the risk greater than the reward. The theory of deterrence was first proposed by Italian philosopher and economist Cesare Beccaria in the 1700s \cite{beccaria}. According to Beccaria, there are three factors of a punishment that determine if the punishment will deter criminal action. These factors are severity, celerity, and certainty.

Since then, the different weights of these factors, and their impact on deterrence at all, have been questioned \cite{deterrence}. Research has shown that the severity of punishment might have the opposite effect; offenders that have faced strict punishment are often more likely to reoffend . The swiftness of punishment has been observed to have little effect at all \cite{uscourts}. Certainty, however, does have a measurable deterrent effect. This has been proven in multiple studies across the world, both in macro- and micro-level contexts \cite{deterrencejustice}.

A more recent theory that has grown out of the idea of deterrence is crime prevention. There are three main forms of crime prevention that have been agreed upon within the field: situational crime prevention, developmental crime prevention, and community crime prevention. Situational crime prevention, or SCP, focuses on proximate causes of crime. As the name suggests, SCP focuses on how to decrease a specific crime within a specific context. Its goal is to prevent crime by eliminating or reducing conditions that make a criminal offense more likely to occur. It draws heavily on economic theory about decision making \cite{criminology}.

In 2003, two Criminal Justice scholars Cornish and Clarke identified 25 specific techniques of crime prevention. Many of these techniques directly conflict with privacy, specifically those within the "Increase the Risks" category. This category is rather self-explanatory and highlights techniques that increase the risks of performing criminal activities. The two techniques most at odds with privacy are "reduce anonymity" and "strengthen formal surveillance." \cite{crimeprevention}

\section{Privacy Technology}
Privacy technology increases anonymity, and in doing so shrinks the certainty factor of deterrence from cybercrime to almost nothing. In this way, the development of PETs is in direct conflict with crime prevention. One of the most famous PETs is Tor. An acronym for "The Onion Router," Tor is a service that sends messages with layers of encryption between multiple endpoints to obfuscate network traffic and provide anonymity for users. According to RSA, the group that utilizes Tor the most is cybercriminals. Tor’s criminal use cases include: the trade of stolen financial data, financial fraud, illegal sexual content, bypassing censorship, drug trafficking, weapons trading, gambling, and the sale of stolen goods \cite{cybercrime}.

Tor is also a popular area of security research. One of the 2019 PoPETs papers was, "DPSelect: A Differential Privacy Based Guard Relay Selection Algorithm for Tor" \cite{dpselect}. Researchers have recently identified a system-wide vulnerability of Tor against network-level adversaries, which includes governments and other large policing bodies. These adversaries can analyze traffic to glean information about users. In addition to passive attacks, researchers have found that Tor users can be de-anonymized using active attacks that target BGP. There has been a proposed counter to this particular attack, called Counter-RAPTOR. Counter-RAPTOR uses a "guard relay selection algorithm" to select guard relays with a high resilience to BGP hijacking attacks. In Tor, guard relays are the first "messengers" in the relay system, taking a message from the user’s IP, packaging it, and sending it to another layer. As a result of having access to IPs, guard relays have to be more trusted and resilient than other nodes in the Tor ecosystem. 

Hanley et al propose a new guard relay selection algorithm called DPSelect that uses differential privacy to improve upon Counter-RAPTOR, which can leak information through the decrease in randomness of guard relay selection because it is location based. Over time, an adversary could start seeing patterns in which guard relays are selected, and learn more about the Tor users. DPSelect mitigates this by including a "Max-Divergence" metric to adjust the likelihood of a particular potential guard relay being chosen. Max-Divergence is equal to the natural log of the highest probability of choosing a specific guard relay over the lowest probability of picking that relay. In a truly random scenario, each relay has an equal chance of being picked, so you get ln(1), which is equal to 0. By setting the Max-Divergence, the algorithm roughly determines how far from truly random you are willing to go, which determines how much information could be leaked. A higher Max-Divergence means a higher deviation from random. The Max-Divergence of DPSelect was 0.67 in the average case and 1.05 in the worst case, compared to Counter-RAPTOR with 1.3 in the average case and 7 in the worst case. Ultimately, this means that the new DPSelect algorithm reduces the ability for network-level actors to de-identify Tor users.

Another paper looks at the flaws of Tor and proposes an entire alternative service in, ConsenSGX: Scaling Anonymous Communications Networks with Trusted Execution Environments"\cite{consensgx}. In their design, ORAM protocols are utilized to fetch smaller portions of the full network without revealing the address space they learned about so as to minimize an adversary’s ability to attack the network. ORAM stands for Oblivious RAM; it’s a cryptographic tool to prevent the exposure of information to anyone observing memory access patterns. This team set out to design a network that is scalable, efficient, and requires minimal changes to the underlying Tor architecture. 

On the server side, they built their network protocols on top of Tor’s. The main difference is how ConsenSGX does directory authorities. Each directory authority distributes its own parameters. These parameters contain all things that are not for an individual relay like protocol version, network features, and the number of relays in the different pools (as well as the bandwidth for these pools.) The directory authorities also need to verify the caches that serve the protocol. Each of the DAs has a long TTL signature verification key that is then used to sign an ephemeral asymmetric keypair. This key is then used to verify what the protocol calls an attestation. Once verified, the server responds with descriptions of available relays. This process can only work if a client knows if a node supports ConsenSGX.On the client side, like the Tor protocol, they connect to a DA authenticating and verify using the DA’s public key. Once authenticated to the DA, the client selects a relay to connect to.

In the evaluation of the ConsenSGX, the team found that the protocol is faster than it’s PIR-Tor counterpart. This speed increase is because the time complexity of client queries in ConsenSGX is log3(N), whereas Tor is O(N). Beyond the time complexity benefits, The amount of bandwidth required is also less. These benefits makes it easy to deploy a ConsenSGX node. The ConsenSGX scheme is built on top of Intel SGX as its Trusted Execution Environments. This scheme was shown to be an effective implementation for scalable anonymous networks.

Anonymous communication is another tool that can help enable cybercrime. A new covert channel called Tithonus was developed by Recabarren and Carbunar and described in their paper, "Tithonus: A Bitcoin Based Censorship Resilient System"\cite{tithonus}. It is specifically designed to not be able to be brought down by state level actors, which are typically the ones trying to catch and prevent illegal activities. Tithonus uses Bitcoin’s gossip protocol rather than relying on full consensus. The authors try to provide 6 properties. First is unobservability, which guarantees that a censor would be unable to detect communications even if they can inspect packets and corresponding metadata. This is measured in both unobservable access to received messages as well as sent messages that are indistinguishable from normal Bitcoin usage. Next, unblockability says that the censor is unable or unwilling to block communications even if the unobservability property was unable to be achieved because that would mean disrupting normal function of the blockchain. Availability promises that the system is resilient to DOS attacks. Tithonus hampers DOS attacks by requiring that requests be paid upfront, so that all interaction with the system forces the users to invest resources. Communication integrity says that communications between the user and the destination should not be able to be modified. The property of ease of deployment is fairly self explanatory. Along with being easy to bootstrap and deploy, Tithonus does not require altruistic participation (when clients download blockchain content and persist it with no remuneration). Finally, performance promises that the system minimizes cost while maximizing the amount of useful information sent over time.

Tithonus utilizes a complex communication stack to achieve its goals. The lowest layer embeds data into transactions, the next optimizes transactions fees, the one after that sends messages of random size, the next establishes trust, the layer on top of that allows the client to securely communicate with the system, and the final layer is the application layer, providing the interface. Based on the ease of deployment property, users don’t need to understand or be aware of the intricacies of the implementation. Tithonus even limits the amount of content a client can request per day to be consistent with regular Bitcoin users which provides further obscurity for users. 

\section{Conclusion}
According to Criminal Justice theory that has been developed and tested over centuries, the more certain someone is that they can commit a crime and get away with it, the more likely they are to do so. The developments in Privacy Enhancing Technologies have only made it easier to commit a crime undetected. PETs prevent governments and other policing bodies from being able to detect, monitor, track, and even prove criminal activity. Improvements in Tor further prevent users from being identified. Alternatives to Tor allow the technology to take root at a much larger scale. Breakthroughs in covert communication channels facilitate the planning and conducting of criminal activity. Disabling the ability to combat cybercrime is an important ethical concern. While preserving privacy is a noble goal, researchers must consider how their developments might be used.

\end{document}